%% file: 2027_isac_arxiv.tex
\documentclass[conference]{IEEEtran}

\usepackage{amsmath,amssymb,amsfonts}
\usepackage{graphicx}
\usepackage{booktabs}
\usepackage{cite}
\usepackage{xcolor}
\usepackage{placeins}
\usepackage[export]{adjustbox}

\newcommand{\rr}{\mathbf r}
\newcommand{\vv}{\mathbf v}
\newcommand{\ttx}{\mathbf t}
\newcommand{\srx}{\mathbf s}
\newcommand{\ut}{\mathbf u_{\mathrm t}}

\newcommand{\bl}{\mathbf b}
\newcommand{\J}{\mathbf J}
\newcommand{\Jeq}{\mathbf J_{\mathrm{eq}}}

\newcommand{\blfootnote}[1]{%
  \begingroup
  \renewcommand{\thefootnote}{}\footnote{#1}%
  \addtocounter{footnote}{-1}%
  \endgroup
}

\title{Bistatic Delay-Doppler Sensing for 3D Target Localization in Distributed Wi-Fi Systems}

\newcommand{\emailat}{@}
\newcommand{\us}{\_}

\author{
\IEEEauthorblockN{
Kyungmin Hong,
Moein Ahmadi,
Thomas Stifter\textsuperscript{*},
Mounim A. El Yacoubi\textsuperscript{\ensuremath{\dagger}},
Bhavani Shankar Mysore R
}
\IEEEauthorblockA{
\textit{Interdisciplinary Centre for Security, Reliability and Trust (SnT), University of Luxembourg, Luxembourg}\\
\textit{\textsuperscript{*}IEE S.A., Luxembourg; \textsuperscript{\ensuremath{\dagger}}T{\'e}l{\'e}com SudParis, Institut Polytechnique de Paris, France}\\
\textit{e-mails: \{kyungmin.hong, moein.ahmadi, bhavani.shankar\}\emailat{}uni.lu, thomas.stifter\emailat{}iee.lu}\\
\textit{mounim.el\us{}yacoubi\emailat{}telecom-sudparis.eu}
}
}

\begin{document}
\maketitle

\blfootnote{%
Accepted for presentation at the 1st IEEE Conference on Integrated Sensing
and Communications (ISAC), Lisbon, Portugal, 15--18 November 2026. This is
the authors' accepted version; the final version of record will appear in
IEEE Xplore.
\copyright~2026 IEEE. Personal use of this material is permitted. Permission
from IEEE must be obtained for all other uses, in any current or future
media, including reprinting/republishing this material for advertising or
promotional purposes, creating new collective works, for resale or
redistribution to servers or lists, or reuse of any copyrighted component of
this work in other works.%
}

\begin{abstract}
This paper investigates how bistatic delay information improves the observability and localization performance of 3D moving targets in distributed Wi-Fi sensing. Motivated by emerging IEEE 802.11bf WLAN sensing, we consider one transmitter, multiple distributed receivers, and a single target with unknown position and velocity. We compare Doppler-only sensing with delay-Doppler sensing, in which delay-capable receivers provide additional bistatic path-length information for target localization. We derive the Fisher information matrix (FIM) and the position equivalent FIM (EFIM) for both sensing formulations. The resulting observability analysis shows that Doppler-only localization generically requires six receivers, whereas delay-Doppler localization can become locally identifiable with three receivers under non-degenerate geometry. We further study mixed receiver architectures in which only a subset of receivers provides delay estimates. Simulations with Wi-Fi-like bandwidths confirm the transition and quantify the roles of bandwidth, Doppler noise, receiver geometry, and delay-capable receiver availability.
\end{abstract}

\begin{IEEEkeywords}
ISAC, delay-Doppler sensing, Fisher information, EFIM, CRB, receiver configuration, Wi-Fi sensing.
\end{IEEEkeywords}

\section{Introduction}

Integrated sensing and communication (ISAC) is being developed as a core 6G capability because communication infrastructure can also provide environmental sensing and localization services \cite{delima2021convergent_6g,liu2020joint_radar_comm_tcom,liu2022isac_jsac,gonzalezprelcic2024isac_revolution}. Recent surveys describe the signal-processing stack for joint communication and radar sensing \cite{zhang2021jcr_jstsp,zhang2022jcr_mobile_survey} and the evolution of ISAC signal designs toward 5G-A and 6G \cite{wei2023isacsignals_iotj}. Millimeter-wave joint radar-communication studies have investigated signal-processing frameworks and system designs for this integration \cite{8828030,8642926}. Moreover, delay-Doppler representations have become important because delay and Doppler directly encode geometric propagation parameters relevant to both communication and radar sensing \cite{mohammed2022otfs_bits}.

Wi-Fi is a particularly relevant ISAC platform because IEEE 802.11 devices are already widely deployed indoors and recent WLAN-sensing efforts, including IEEE 802.11bf, explicitly target sensing functions such as presence detection, tracking, and localization using 802.11-compliant transmissions \cite{restuccia2021wifi_sensing,meneghello2022wifi_isac,du2022wlan_sensing,sahoo2024wifi_bf_protocol}. In such systems, channel measurements obtained from Wi-Fi packets can be processed to estimate geometric quantities such as delay and Doppler, while the sensing algorithm itself remains implementation-dependent.

Prior work investigated Doppler-based 3D localization using distributed bistatic Wi-Fi links and an RF digital twin \cite{Hong2026IRS}. Here, we address a complementary measurement-level CRB question. In 3D localization, each Doppler measurement depends on both the target position and the unknown velocity; treating the velocity as a nuisance parameter reduces the effective position information and raises the number of independent receiver observations required for a nonsingular position bound. We then add bistatic delay measurements, expressed through propagation path length, which constrain the target position directly while the Doppler measurements retain the motion information. The central question is therefore whether adding bistatic delay can reduce the receiver-count burden of Doppler-only 3D localization, and under what conditions this benefit is significant.

The contribution is threefold. First, we formulate Doppler-only and delay-Doppler FIMs for a single moving target observed by one Wi-Fi-like transmitter and multiple distributed receivers. Second, using the position equivalent Fisher information matrix (EFIM) methodology for localization bounds \cite{shen2010wideband_part1,shen2010wideband_part2}, we derive receiver-count conditions after eliminating the unknown velocity and extend the analysis to mixed delay-capable receiver architectures. Third, we support the receiver-count transition with position error bound (PEB) statistics, spatial PEB maps, bandwidth and Doppler-noise sensitivity studies, mixed-architecture heatmaps, a delay-only baseline that separates the delay and Doppler contributions, and a local nonlinear least-squares (NLS) consistency check.

\section{System Model}
We consider a sub-7 GHz Wi-Fi bistatic sensing setup with a transmitter at $\ttx\in\mathbb R^3$, receivers at $\srx_\ell\in\mathbb R^3$, $\ell=1,\ldots,L$, and a moving target with position $\rr\in\mathbb R^3$ and velocity $\vv\in\mathbb R^3$. Each receiver is assumed to provide delay and/or Doppler measurements for the target path. The unknown target parameter vector is
\begin{equation}
\boldsymbol\theta = [\rr^T,\vv^T]^T.
\end{equation}

\begin{figure}[t]
\centering
\includegraphics[width=0.90\columnwidth]{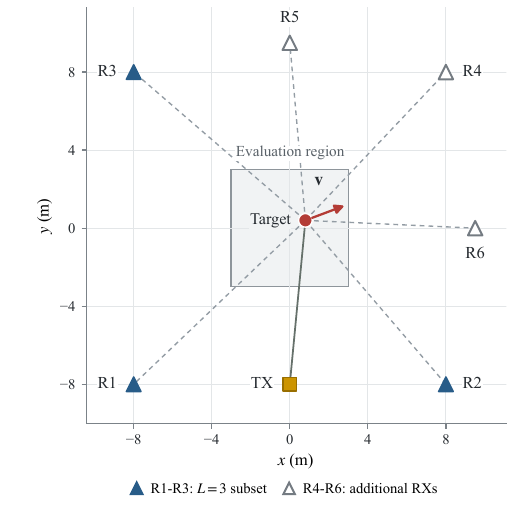}
\caption{Top-view distributed bistatic Wi-Fi sensing geometry. Blue receivers indicate the $L=3$ delay-Doppler subset, while gray receivers indicate additional candidate receiver locations used in the sweep up to $L=6$. The deployment is not coplanar: the transmitter is at height $2.5$ m and the six candidate receivers span heights $1.2$--$3.6$ m.}
\label{fig:geometry}
\end{figure}
We will also define the following for use in the sequel
\begin{equation}
\ut(\rr)=\frac{\rr-\ttx}{\|\rr-\ttx\|},\quad
\mathbf u_{\mathrm r,\ell}(\rr)=\frac{\rr-\srx_\ell}{\|\rr-\srx_\ell\|},
\end{equation}
and
\begin{equation}
\bl_\ell(\rr)=\ut(\rr)+\mathbf u_{\mathrm r,\ell}(\rr).
\end{equation}
\subsection{Measurement models}

We consider a Doppler-only baseline and a delay-Doppler model, a comparison that isolates the geometric information added by bistatic delay.

\subsubsection{Doppler-Only Baseline}
The Doppler-only measurement at receiver $\ell$ is
\begin{equation}
f_{D,\ell}=\alpha \vv^T\bl_\ell(\rr)+n_{f,\ell},
\end{equation}
where $\alpha=f_c/c$ and $n_{f,\ell}\sim\mathcal N(0,\sigma_f^2)$.

\subsubsection{Delay-Doppler Model}
The physical bistatic propagation delay is
\begin{equation}
\tau_\ell(\rr)
=
\frac{\rho_\ell(\rr)}{c}
=
\frac{\|\rr-\ttx\|+\|\rr-\srx_\ell\|}{c}.
\end{equation}
Here $\rho_\ell(\rr)$ denotes the total bistatic propagation path length from the transmitter to the target and then to receiver $\ell$. In the FIM analysis, we express the bistatic delay in meters as
\begin{equation}
\tilde\rho_\ell
=
\rho_\ell(\rr)+n_{\rho,\ell},
\quad
\rho_\ell(\rr)
=
c\tau_\ell(\rr)
=
\|\rr-\ttx\|+\|\rr-\srx_\ell\|.
\end{equation}
The noise term $n_{\rho,\ell}\sim\mathcal N(0,\sigma_\rho^2)$ is expressed in meters and represents delay-estimation uncertainty after conversion to bistatic path length. The Doppler measurement remains
\begin{equation}
f_{D,\ell}(\rr,\vv)=\alpha \vv^T\bl_\ell(\rr)+n_{f,\ell}.
\end{equation}

This measurement-level abstraction separates geometric observability from implementation-specific Wi-Fi processing: the bounds below quantify the accuracy achievable for given error variances rather than evaluating one estimator. Two assumptions are implied. First, the bistatic path length is observed without an unknown common clock bias. We assume receivers are time-aligned to the transmitter, for instance through a disciplined reference or by calibrating the direct-path arrival, so that residual offsets are absorbed into $n_{\rho,\ell}$; unmodelled per-receiver clock offsets would add nuisance parameters and raise the receiver count needed for a nonsingular position EFIM. Second, every receiver reports the delay and Doppler of the same target path. Indoor receivers observe many static and dynamic components, so the bounds are conditional on correct detection and cross-receiver association of an unbiased single-target measurement. Association errors and unresolved multipath lie outside the present scope and would degrade accuracy relative to these bounds, which should therefore not be read as end-to-end IEEE 802.11bf performance.

\section{FIM and Position EFIM}
We use a measurement-level Gaussian FIM, consistent with CRB-based localization analyses for wideband networks and distributed radar \cite{shen2010wideband_part1,godrich2010mimo_radar}.

The delay Jacobian is
\begin{equation}
\frac{\partial \rho_\ell}{\partial \rr}
=
\bl_\ell(\rr),
\quad
\frac{\partial \rho_\ell}{\partial \vv}
=
\mathbf 0 .
\end{equation}
For Doppler,
\begin{equation}
\frac{\partial f_{D,\ell}}{\partial \vv}
=
\alpha \bl_\ell(\rr),
\end{equation}
and
\begin{equation}
\frac{\partial f_{D,\ell}}{\partial \rr}
=
\alpha \mathbf G_\ell(\rr)\vv,
\end{equation}
where
\begin{equation}
\mathbf G_\ell(\rr)
=
\frac{\mathbf I-\ut(\rr)\ut^T(\rr)}
{\|\rr-\ttx\|}
+
\frac{
\mathbf I-\mathbf u_{\mathrm r,\ell}(\rr)\mathbf u_{\mathrm r,\ell}^T(\rr)
}
{\|\rr-\srx_\ell\|}.
\end{equation}

For delay-Doppler measurements, define
\begin{equation}
\mathbf q_{\rho,\ell}
=
\begin{bmatrix}
\bl_\ell^T & \mathbf 0_3^T
\end{bmatrix}^T,
\quad
\mathbf q_{f,\ell}
=
\begin{bmatrix}
(\alpha\mathbf G_\ell\vv)^T &
(\alpha\bl_\ell)^T
\end{bmatrix}^T .
\end{equation}
The per-receiver FIM contribution is
\begin{equation}
\J_\ell
=
\frac{1}{\sigma_\rho^2}
\mathbf q_{\rho,\ell}\mathbf q_{\rho,\ell}^T
+
\frac{1}{\sigma_f^2}
\mathbf q_{f,\ell}\mathbf q_{f,\ell}^T .
\end{equation}
The total FIM is
\begin{equation}
\J
=
\sum_{\ell=1}^{L}\J_\ell .
\end{equation}
For the Doppler-only baseline, the delay observations are excluded, so only the Doppler term of $\J_\ell$, proportional to $1/\sigma_f^2$, is retained.

We also consider a mixed architecture in which all $L$ receivers provide Doppler measurements but only a subset $\mathcal D$ with $|\mathcal D|=K$ provides delay estimates:
\begin{equation}
\J_{\mathrm{mix}}
=
\sum_{\ell=1}^{L}
\frac{1}{\sigma_f^2}
\mathbf q_{f,\ell}\mathbf q_{f,\ell}^T
+
\sum_{\ell\in\mathcal D}
\frac{1}{\sigma_\rho^2}
\mathbf q_{\rho,\ell}\mathbf q_{\rho,\ell}^T .
\end{equation}
This model reduces to Doppler-only sensing when $K=0$ and to full delay-Doppler sensing when $K=L$.
The position EFIM after eliminating the unknown velocity is
\begin{eqnarray}
\Jeq
&=&
\J_{\rr\rr}
-
\J_{\rr\vv}
\J_{\vv\vv}^{\dagger}
\J_{\vv\rr}, ~~{\rm where} \\
\J
&=&
\begin{bmatrix}
\J_{\rr\rr} & \J_{\rr\vv}\\
\J_{\vv\rr} & \J_{\vv\vv}
\end{bmatrix}.
\end{eqnarray}

For nonsingular $\Jeq$, the position error bound (PEB) is
\begin{equation}
\mathrm{PEB}
=
\sqrt{
\mathrm{tr}
\left(
\Jeq^{-1}
\right)
},
\end{equation}
i.e. $\mathbb E\{\|\hat\rr-\rr\|^2\}\geq\mathrm{PEB}^2$ for every unbiased estimator of $\rr$ with $\vv$ unknown. It is a lower bound on the root-mean-square 3D position error under the assumed measurement model; attainment is not implied. It is reported in metres throughout.

\section{Receiver-Count Observability Analysis}
Stack the measurement Jacobians as $\mathbf H=[\mathbf H_{\rr}~~\mathbf H_{\vv}]$, one row per scalar measurement, with noise covariance $\boldsymbol\Sigma$, and let $\mathbf P$ be the orthogonal projector onto the complement of $\mathrm{range}(\boldsymbol\Sigma^{-1/2}\mathbf H_{\vv})$. Eliminating the velocity is exactly this projection, since $\Jeq=(\mathbf P\boldsymbol\Sigma^{-1/2}\mathbf H_{\rr})^T(\mathbf P\boldsymbol\Sigma^{-1/2}\mathbf H_{\rr})$, so that
\begin{equation}
\label{eq:rankcond}
\Jeq\succ\mathbf 0
\iff
\mathrm{rank}\!\left(\mathbf P\boldsymbol\Sigma^{-1/2}\mathbf H_{\rr}\right)=3 .
\end{equation}
Delay rows satisfy $\partial\rho_\ell/\partial\vv=\mathbf 0$ and are untouched by the projection, whereas the $L$ Doppler rows lose three dimensions to the unknown velocity. With $K$ delay-capable receivers, the number of position-informative rows that survive is $K+\max(L-3,0)$, so a necessary condition for \eqref{eq:rankcond} is
\begin{equation}
\label{eq:count}
K+\max(L-3,0)\geq3 ,
\end{equation}
which for $L\geq3$ is exactly $L+K\geq6$. It specializes to $L\geq6$ for Doppler-only sensing ($K=0$) and to $L\geq3$ for delay-Doppler sensing ($K=L$).

One special case is worth isolating. When $L\leq3$ and the Doppler velocity Jacobian has full row rank, the three unknown velocity components absorb every Doppler row, so the projection annihilates them and
\begin{equation}
\label{eq:l3equiv}
\Jeq=\frac{1}{\sigma_\rho^2}\sum_{\ell=1}^{L}\bl_\ell\bl_\ell^T ,
\end{equation}
exactly the delay-only position FIM. At $L=3$, delay-Doppler and delay-only sensing therefore yield identical position bounds: the identifiability recovered at three receivers comes entirely from the delay measurements. Doppler can contribute additional position information once $L>3$, although that contribution vanishes again for a stationary target.

Four statements must be distinguished. (i) Under the generic full-column-rank assumption for the Doppler velocity Jacobian, condition \eqref{eq:count} is \emph{necessary}. (ii) It is \emph{generically sufficient} for a configuration whose projected Jacobian admits a full-rank realization: the entries of $\mathbf H$ are real-analytic in $(\rr,\vv,\{\srx_\ell\})$ away from degenerate placements, so a single non-vanishing $3\times3$ minor of $\mathbf P\boldsymbol\Sigma^{-1/2}\mathbf H_{\rr}$ forces rank deficiency onto a set of Lebesgue measure zero. Every configuration reported as identifiable in Section~V exhibits such realizations. Examples of rank-deficient cases include a stationary target under Doppler-only sensing and a coplanar transmitter-receiver-target geometry with in-plane target velocity. (iii) For the layout of Fig.~\ref{fig:geometry} the transition is \emph{verified numerically} in Section~V. (iv) We do \emph{not} claim a guarantee for every non-degenerate geometry, and a full-rank $\Jeq$ does not imply small error: with large delay uncertainty, limited bandwidth, or poor conditioning, delay measurements may restore rank while the PEB remains large.

\section{Numerical Results}
\input{table_sim_params}
Statistics are computed over $2000$ target states per configuration, sampled with $x,y\sim\mathcal U(-3,3)$ m, $z\sim\mathcal U(1,3)$ m, speed $\|\vv\|\sim\mathcal U(0.3,1.5)$ m/s and isotropic direction. The receiver-count sweep draws an independent state set for each configuration, whereas the mixed-architecture sweep reuses one common state set across all $(L,K)$, so those comparisons are paired. The same receiver ordering is used for all receiver-count sweeps, so $L=3$ corresponds to the blue subset in Fig.~\ref{fig:geometry}. Unless otherwise stated, PEB statistics are computed from the position EFIM after eliminating the unknown velocity.

A trial is declared invalid when the position EFIM $\Jeq$ is not numerically positive definite; in the implementation a trial is valid only if all eigenvalues of $\Jeq$ are finite and $\lambda_{\min}(\Jeq)>10^{-9}$ m$^{-2}$. Although this is an absolute threshold, the two regimes are far apart: rank-deficient configurations return $\lambda_{\min}$ at round-off level (median $|\lambda_{\min}|\approx10^{-12}$ m$^{-2}$), whereas full-rank configurations have median $\lambda_{\min}$ of order $10^{-2}$ to $10^{1}$ m$^{-2}$. Any threshold between $10^{-12}$ and $10^{-6}$ m$^{-2}$ changes the invalid fractions in Table~\ref{tab:key_results} by at most $0.5$ percentage points, so the classification is not an artefact of this choice. The EFIM is formed by projecting the position Jacobian off the range of the velocity Jacobian rather than by evaluating $\J_{\rr\vv}\J_{\vv\vv}^{\dagger}\J_{\vv\rr}$ explicitly; the two are algebraically identical, but the explicit difference loses small eigenvalues to cancellation when $\J_{\vv\vv}$ is ill-conditioned. Reported PEB statistics are computed over valid trials only, while the invalid fraction is reported separately. For the local NLS check the estimator is initialized around the true state; it probes local measurement-noise behaviour and local identifiability only, and is not proposed as a usable localization algorithm.

\subsection{Receiver-Count Transition}

\begin{figure}[!tbp]
\centering
\begin{minipage}{0.48\columnwidth}
\centering
\includegraphics[width=\linewidth]{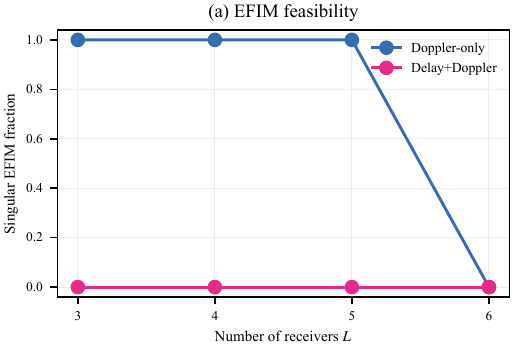}
\end{minipage}\hfill
\begin{minipage}{0.48\columnwidth}
\centering
\includegraphics[width=\linewidth]{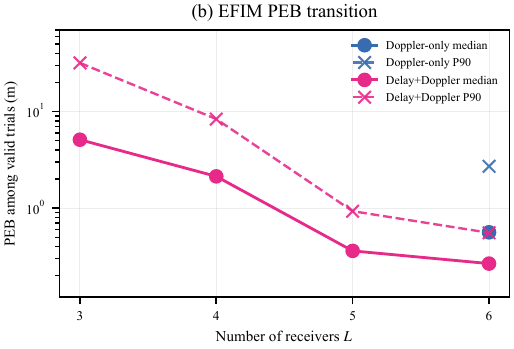}
\end{minipage}\\[0.65em]
\begin{minipage}{0.55\columnwidth}
\centering
\includegraphics[width=\linewidth]{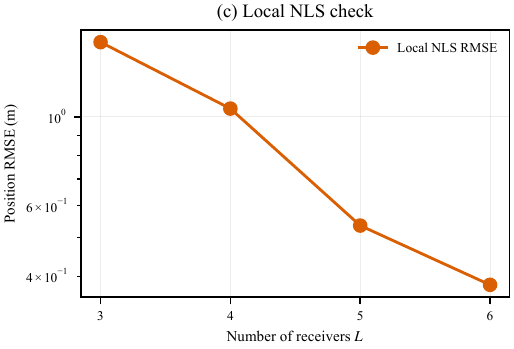}
\end{minipage}
\caption{Receiver-count transition and local NLS consistency check. 
(a) EFIM singularity fraction versus receiver count. 
(b) EFIM-based median and P90 PEB among valid trials. 
(c) Empirical local NLS RMSE at $B=80$ MHz.}
\label{fig:receiver_count}
\end{figure}

Fig.~\ref{fig:receiver_count} reports the main receiver-count transition. Panel (a) shows the EFIM singularity fraction: the Doppler-only EFIM stays singular until the receiver set is large enough, while the delay-Doppler EFIM is already nonsingular at $L=3$ for the tested non-degenerate layout. Panel (b) shows that the PEB quantiles among valid trials improve as receivers are added. Panel (c) is a local NLS consistency check at $B=80$ MHz, initialized near the true state to avoid global initialization failures, and used only to verify that the empirical error decreases with receiver count.

The $L=1,2$ cases are omitted because their position EFIMs are rank-deficient in this 3D setup. Panel (c) does not claim CRB efficiency; it only confirms that the local NLS error follows the same trend as the EFIM analysis.

\subsection{Spatial PEB Behavior}

\begin{figure}[!tbp]
\centering
\includegraphics[width=0.92\columnwidth]{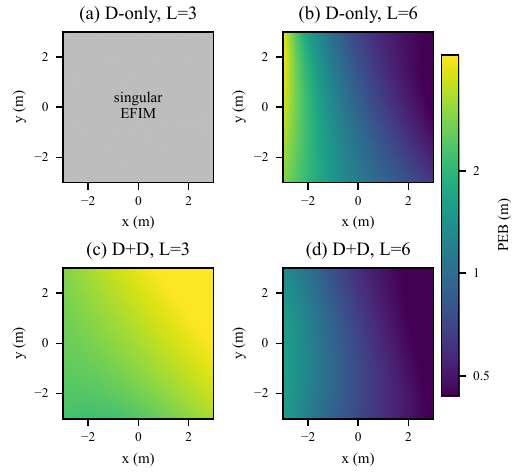}
\caption{Spatial PEB maps on the horizontal target-position slice $z=1.6$ m for the geometry in Fig.~\ref{fig:geometry}. Panels labelled D-only use Doppler-only sensing; panels labelled D+D use delay-Doppler sensing, where the first D stands for delay. Gray regions indicate singular or numerically invalid EFIM instances.}
\label{fig:peb_maps}
\end{figure}

The receiver-count transition is not uniform over the target region. Fig.~\ref{fig:peb_maps} maps the effect of adding delay information on a horizontal slice: the Doppler-only $L=3$ case is rank-deficient throughout, whereas delay-Doppler sensing yields finite bounds at the same receiver count, and $L=6$ further reduces the PEB and smooths its spatial variation.

\subsection{Mixed Delay-Capable Architectures}

\begin{figure}[!tbp]
\centering
\adjustbox{trim=0 0 {.50\width} 0,clip,width=0.94\columnwidth}{\includegraphics{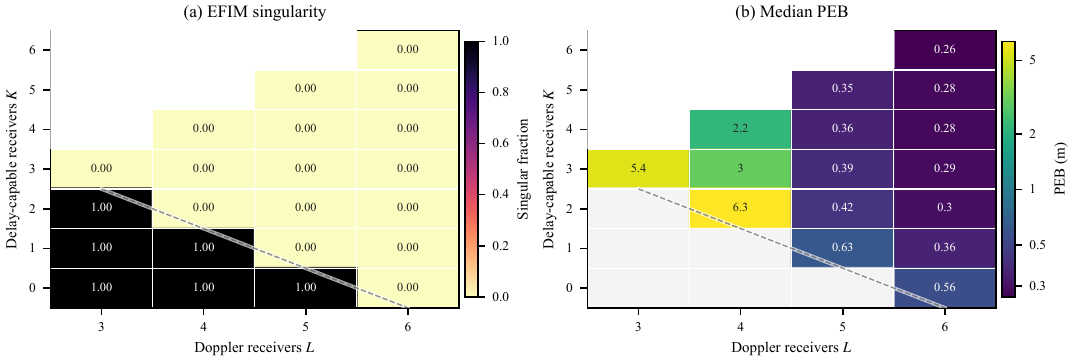}}\\[-0.25em]
\adjustbox{trim={.50\width} 0 0 0,clip,width=0.94\columnwidth}{\includegraphics{figures/fig06_mixed_delay_capable_heatmap.pdf}}
\caption{Mixed receiver architecture where all $L$ receivers provide Doppler measurements and only $K$ receivers provide delay estimates. The $K$ delay-capable receivers are the first $K$ of the fixed ordering in Fig.~\ref{fig:geometry}. Blank cells are the infeasible combinations with $K>L$; in the PEB map they also include the combinations whose position EFIM is singular. The dashed line marks the boundary $L+K=6$ of \eqref{eq:count}. The singularity map shows rank recovery, while the median-PEB map shows that finite accuracy still depends on geometry and delay-capable receiver availability.}
\label{fig:mixed_architecture}
\end{figure}

Fig.~\ref{fig:mixed_architecture} extends the receiver-count transition to mixed delay-capable deployments. In this experiment, all $L$ receivers provide Doppler measurements, while only $K$ of them provide delay estimates. The $K$ delay-capable receivers are the first $K$ in the fixed ordering of Fig.~\ref{fig:geometry}, so the sweep varies $K$ at a fixed geometry rather than optimizing the subset. The choice is not neutral: enumerating all $\binom{6}{K}$ subsets at $L=6$ spans best-to-worst median PEBs of $0.35$--$0.55$ m for $K=1$, $0.30$--$0.50$ m for $K=2$ and $0.27$--$0.45$ m for $K=3$, and the ordering used here ranks first, first and third respectively, so the heatmap is close to the best case. Delay capability is therefore most useful on receivers whose bistatic geometry complements the Doppler-only set. The Doppler-only cases with $L<6$ remain singular, whereas adding a small number of delay-capable links can recover a nonsingular position EFIM, following the boundary $L+K=6$ of \eqref{eq:count}; the PEB map also shows that rank recovery and high accuracy are distinct requirements.

For example, $(L,K)=(5,1)$ has a median PEB of about $0.63$ m, close to the Doppler-only $(6,0)$ case at about $0.56$ m, while adding one delay-capable link to six receivers reduces it to about $0.36$ m. By contrast $(3,3)$ and $(4,2)$ are locally identifiable yet still have multi-metre median PEB. A delay-capable link adds a position constraint, but its value depends on delay accuracy and on bistatic geometry, so $(5,1)$ recovers identifiability without necessarily beating $(6,0)$.

\subsection{Bandwidth and Doppler-Noise Sensitivity}

For the bandwidth-sensitivity study, we model the delay-estimation uncertainty after conversion to bistatic path length as
\begin{equation}
\sigma_\rho(B)
=
\sigma_{\rho,\mathrm{ref}}
\frac{B_{\mathrm{ref}}}{B}.
\end{equation}
This captures the standard trend that larger bandwidth improves delay precision. The projection onto target coordinates additionally depends on the bistatic geometry through $\|\ut+\mathbf u_{\mathrm r,\ell}\|=2\cos(\beta_\ell/2)$, which the delay Jacobian already includes. We use $\sigma_{\rho,\mathrm{ref}}=0.50$ m at $B_{\mathrm{ref}}=80$ MHz, as summarized in Table~\ref{tab:sim_params}. The reference noise levels are motivated by idealized estimation bounds. For Doppler, a single-tone frequency CRB with $T_{\mathrm{obs}}=0.2$ s, $N=100$ samples (an illustrative $500$ packet/s sensing rate) and a per-sample SNR of $10$ dB gives approximately $0.062$ Hz, comparable to $\sigma_f=0.0619$ Hz, equivalent to a bistatic range-rate uncertainty of $3.2$ mm/s at $5.8$ GHz. For delay, an idealized CRB with RMS bandwidth $\beta_{\mathrm{rms}}=B/\sqrt{12}$ and an effective SNR of $10$ dB yields approximately $0.46$ m at $B=80$ MHz, comparable to $\sigma_{\rho,\mathrm{ref}}=0.50$ m. These estimation bounds should not be confused with the nominal resolution scale $c/B=3.75$ m, and they do not represent experimentally validated Wi-Fi sensing accuracy. The carrier frequency and bandwidth values represent sub-7 GHz Wi-Fi sensing with common 802.11 channel widths rather than a dedicated radar waveform.

\begin{figure}[!tbp]
\centering
\includegraphics[width=0.92\columnwidth]{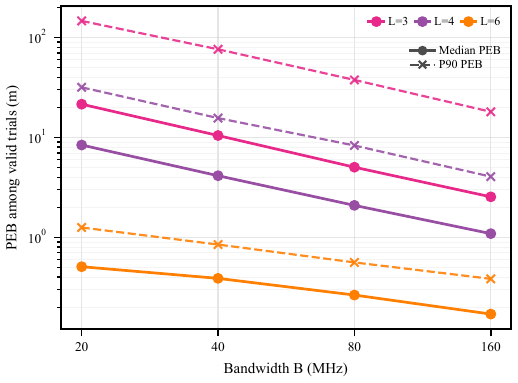}
\caption{Bandwidth sensitivity of delay-Doppler localization. Solid curves show median PEB and dashed curves show P90 PEB among valid trials. Larger bandwidth reduces the delay-estimation uncertainty according to $\sigma_\rho(B)=\sigma_{\rho,\mathrm{ref}}B_{\mathrm{ref}}/B$, leading to lower position bounds.}
\label{fig:bandwidth}
\end{figure}

Fig.~\ref{fig:bandwidth} evaluates the bandwidth dependence of delay-Doppler localization under the path-length uncertainty model above.

As shown in Fig.~\ref{fig:bandwidth}, increasing the bandwidth reduces both the median and P90 PEB for all tested receiver counts. This trend is consistent with the assumed delay-uncertainty model: wider channels reduce the delay-estimation uncertainty after conversion to bistatic path length, thereby increasing the useful position information provided by the delay measurements. At low bandwidth, the EFIM can be nonsingular, but the corresponding PEB can still be large because the delay-estimation uncertainty remains high.

\begin{figure}[!tbp]
\centering
\includegraphics[width=0.92\columnwidth]{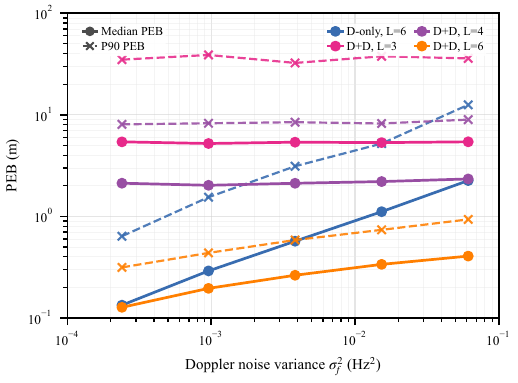}
\caption{Doppler-noise sensitivity. In the legend, D-only denotes Doppler-only sensing and D+D denotes delay-Doppler sensing. Solid curves show median PEB and dashed curves show P90 PEB among valid trials. Increasing $\sigma_f^2$ degrades Doppler-only localization more strongly than delay-Doppler localization, since the delay measurements preserve position information.}
\label{fig:doppler_noise}
\end{figure}

Fig.~\ref{fig:doppler_noise} separates the effect of Doppler precision from the effect of delay information. Doppler-only localization degrades rapidly as $\sigma_f^2$ increases because Doppler is the sole source of information. Delay-Doppler localization is less sensitive to Doppler noise, especially for larger $L$, because the delay measurements continue to constrain the target position.

Receiver count, delay availability, bandwidth and Doppler precision must therefore be considered jointly.

\input{table_key_results}

Table~\ref{tab:key_results} summarizes representative operating points from the receiver-count sweep. The Doppler-only $L=3$ case is rank-deficient for all tested target states, whereas Doppler-only sensing becomes locally feasible near $L=6$. In contrast, delay-Doppler sensing is already locally identifiable at $L=3$, although its median and tail PEB values remain much larger than those obtained with $L=6$. The delay-only rows isolate what Doppler contributes, using the same target states as the delay-Doppler trials. At $L=3$ the two rows are identical in every column, as \eqref{eq:l3equiv} requires. For the sampled velocities, Doppler adds position information from $L=4$ onwards, lowering the median PEB from $2.114$ to $0.266$ m at $L=6$. Neither measurement type dominates on its own: at $L=6$ delay-only ($2.114$ m) is worse than Doppler-only ($0.561$ m), while their combination improves on both.

Two practical implications follow. Doppler-only receivers can eventually recover local identifiability, but the transition needs a relatively large receiver count because the unknown velocity must also be inferred; and a few delay-capable links cut that burden sharply, provided the delay estimates are accurate and the geometry is non-degenerate. The deployment question is therefore not only how many Wi-Fi receivers are available, but which of them can supply reliable delay in addition to Doppler, a distinction that matters in heterogeneous WLAN deployments where links differ in bandwidth, timing resolution, and channel-measurement stability.

\section{Conclusion}
This paper quantified how bistatic delay information changes the local observability of 3D moving-target localization in distributed Wi-Fi sensing. After eliminating the unknown velocity through the position EFIM, Doppler-only sensing generically requires at least six receivers for a full-rank 3D position bound, whereas delay-Doppler sensing can become identifiable with three receivers under non-degenerate geometry. The mixed-architecture study shows that delay-capable links trade off against Doppler receiver count for rank recovery, although rank recovery alone does not guarantee accuracy, and a delay-only baseline separates the two contributions. Delay extraction from Wi-Fi channel measurements is therefore valuable not only for accuracy but for reducing the receiver-count burden of distributed ISAC localization. The analysis is measurement-level and the bounds are conditional: clutter, multiple targets, NLoS bias, unresolved multipath, cross-receiver association errors, and residual synchronization offsets are not modelled, and each would move attainable accuracy away from these bounds. Incorporating them into the EFIM treatment, together with a full 802.11bf processing chain and experimental validation, is left to future work.

\section*{Acknowledgment}
This work was supported by the Luxembourg National Research Fund (FNR) through the ATLAS project (Grant 18060833). The authors thank the project collaborators for their valuable discussions and support.

\bibliographystyle{IEEEtran}
\bibliography{references}

\end{document}

%% file: table_sim_params.tex
\begin{table}[t]
\centering
\caption{Simulation parameters.}
\label{tab:sim_params}
\scriptsize
\begin{tabular}{p{0.39\columnwidth}p{0.53\columnwidth}}
\hline
Parameter & Value \\
\hline
Carrier frequency & $5.8$ GHz \\
Bandwidths & $20,40,80,160$ MHz \\
Reference bandwidth & $B_{\mathrm{ref}}=80$ MHz \\
Delay-noise model &
$\sigma_\rho(B)=\sigma_{\rho,\mathrm{ref}}B_{\mathrm{ref}}/B$ \\
Reference delay std. &
$\sigma_{\rho,\mathrm{ref}}=0.50$ m after conversion to bistatic path length \\
Delay std. &
$\sigma_\tau(B)=\sigma_\rho(B)/c$ \\
Doppler noise std. & $\sigma_f=0.0619$ Hz \\
Receiver counts & $L=3,4,5,6$ \\
\hline
\end{tabular}
\end{table}

%% file: table_key_results.tex
\begin{table}[t]
\centering
\caption{Key receiver-count comparison. ``Invalid frac.'' is the fraction of trials whose position EFIM fails the validity test $\lambda_{\min}(\Jeq)>10^{-9}$ m$^{-2}$; ``Cov.\ 2 m'' is the fraction of valid trials with $\mathrm{PEB}\leq2$ m, undefined when no trial is valid.}
\label{tab:key_results}
\footnotesize
\setlength{\tabcolsep}{3.2pt}
\begin{tabular*}{\columnwidth}{@{\extracolsep{\fill}}lccccc@{}}
\hline
Model & $L$ & Med. & P90 & Invalid & Cov. \\
      &     & PEB (m) & PEB (m) & frac. & 2 m \\
\hline
Doppler-only & 3 & -- & -- & 1.0000 & -- \\
Doppler-only & 6 & 0.561 & 2.712 & 0.0000 & 0.8610 \\
Delay-only & 3 & 5.113 & 31.96 & 0.0000 & 0.0435 \\
Delay-only & 6 & 2.114 & 2.968 & 0.0000 & 0.4460 \\
Delay+Doppler & 3 & 5.113 & 31.96 & 0.0000 & 0.0435 \\
Delay+Doppler & 6 & 0.266 & 0.555 & 0.0000 & 0.9990 \\
\hline
\end{tabular*}
\end{table}

%% file: references.bib
@article{liu2022isac_jsac,
  author={Liu, Fan and Cui, Yuanhao and Masouros, Christos and Xu, Jie and Han, Tony Xiao and Eldar, Yonina C. and Buzzi, Stefano},
  title={Integrated Sensing and Communications: Toward Dual-Functional Wireless Networks for {6G} and Beyond},
  journal={IEEE Journal on Selected Areas in Communications},
  volume={40},
  number={6},
  pages={1728--1767},
  year={2022},
  doi={10.1109/JSAC.2022.3156632}
}

@article{liu2020joint_radar_comm_tcom,
  author={Liu, Fan and Masouros, Christos and Petropulu, Athina P. and Griffiths, Hugh and Hanzo, Lajos},
  title={Joint Radar and Communication Design: Applications, State-of-the-Art, and the Road Ahead},
  journal={IEEE Transactions on Communications},
  volume={68},
  number={6},
  pages={3834--3862},
  year={2020},
  doi={10.1109/TCOMM.2020.2973976}
}

@article{zhang2021jcr_jstsp,
  author={Zhang, J. Andrew and Liu, Fan and Masouros, Christos and Heath, Robert W. and Feng, Zhiyong and Zheng, Le and Petropulu, Athina},
  title={An Overview of Signal Processing Techniques for Joint Communication and Radar Sensing},
  journal={IEEE Journal of Selected Topics in Signal Processing},
  volume={15},
  number={6},
  pages={1295--1315},
  year={2021},
  doi={10.1109/JSTSP.2021.3113120}
}

@article{zhang2022jcr_mobile_survey,
  author={Zhang, J. Andrew and Rahman, Md. Lushanur and Wu, Kai and Huang, Xiaojing and Guo, Y. Jay and Chen, Shanzhi and Yuan, Jinhong},
  title={Enabling Joint Communication and Radar Sensing in Mobile Networks---A Survey},
  journal={IEEE Communications Surveys \& Tutorials},
  volume={24},
  number={1},
  pages={306--345},
  year={2022},
  doi={10.1109/COMST.2021.3122519}
}

@article{wei2023isacsignals_iotj,
  author={Wei, Zhiqing and Qu, Hanyang and Wang, Yuan and Yuan, Xin and Wu, Huici and Du, Ying and Han, Kaifeng and Zhang, Ning and Feng, Zhiyong},
  title={Integrated Sensing and Communication Signals Toward {5G-A} and {6G}: A Survey},
  journal={IEEE Internet of Things Journal},
  volume={10},
  number={13},
  pages={11068--11092},
  year={2023},
  doi={10.1109/JIOT.2023.3235618}
}

@article{mohammed2022otfs_bits,
  author={Mohammed, Saif Khan and Hadani, Ronny and Chockalingam, Ananthanarayanan and Calderbank, Robert},
  title={{OTFS}---A Mathematical Foundation for Communication and Radar Sensing in the Delay-Doppler Domain},
  journal={IEEE BITS the Information Theory Magazine},
  volume={2},
  number={2},
  pages={36--55},
  year={2022},
  doi={10.1109/MBITS.2022.3216536}
}

@article{delima2021convergent_6g,
  author={De Lima, Carlos and Belot, Didier and Berkvens, Rafael and Bourdoux, Andre and Dardari, Davide and Guillaud, Maxime and Isomursu, Minna and Lohan, Elena-Simona and Miao, Yang and Noll Barreto, Andre and Kahar Aziz, Muhammad Reza and Saloranta, Jani and Sanguanpuak, Tachporn and Sarieddeen, Hadi and Seco-Granados, Gonzalo and Suutala, Jaakko and Svensson, Tommy and Valkama, Mikko and Van Liempd, Barend and Wymeersch, Henk},
  title={Convergent Communication, Sensing and Localization in {6G} Systems: An Overview of Technologies, Opportunities and Challenges},
  journal={IEEE Access},
  volume={9},
  pages={26902--26925},
  year={2021},
  doi={10.1109/ACCESS.2021.3053486}
}

@article{gonzalezprelcic2024isac_revolution,
  author={Gonz{\'a}lez-Prelcic, Nuria and Keskin, Musa Furkan and Kaltiokallio, Ossi and Valkama, Mikko and Dardari, Davide and Shen, Xiao and Shen, Yuan and Bayraktar, Murat and Wymeersch, Henk},
  title={The Integrated Sensing and Communication Revolution for {6G}: Vision, Techniques, and Applications},
  journal={Proceedings of the IEEE},
  volume={112},
  number={7},
  pages={676--723},
  year={2024},
  doi={10.1109/JPROC.2024.3397609}
}

@article{restuccia2021wifi_sensing,
  author={Wu, Chenshu and Wang, Beibei and Au, Oscar C. and Liu, K. J. Ray},
  title={{Wi-Fi} Can Do More: Toward Ubiquitous Wireless Sensing},
  journal={IEEE Communications Standards Magazine},
  volume={6},
  number={2},
  pages={42--49},
  year={2022},
  doi={10.1109/MCOMSTD.0001.2100111}
}

@article{meneghello2022wifi_isac,
  author={Meneghello, Francesca and Chen, Cheng and Cordeiro, Carlos and Restuccia, Francesco},
  title={Toward Integrated Sensing and Communications in {IEEE} 802.11bf {Wi-Fi} Networks},
  journal={IEEE Communications Magazine},
  volume={61},
  number={7},
  pages={128--133},
  year={2023},
  doi={10.1109/MCOM.001.2200806}
}

@article{du2022wlan_sensing,
  author={Du, Rui and Hua, Haocheng and Xie, Hailiang and Song, Xianxin and Lyu, Zhonghao and Hu, Mengshi and Narengerile and Xin, Yan and McCann, Stephen and Montemurro, Michael and Han, Tony Xiao and Xu, Jie},
  title={An Overview on {IEEE} 802.11bf: {WLAN} Sensing},
  journal={IEEE Communications Surveys \& Tutorials},
  volume={27},
  number={1},
  pages={184--217},
  year={2025},
  doi={10.1109/COMST.2024.3408899}
}

@inproceedings{sahoo2024wifi_bf_protocol,
  author={Sahoo, Anirudha and Ropitault, Tanguy and Blandino, Steve and Golmie, Nada},
  title={Sensing Performance of the {IEEE} 802.11bf Protocol and Its Impact on Data Communication},
  booktitle={2024 IEEE 100th Vehicular Technology Conference (VTC2024-Fall)},
  pages={1--7},
  year={2024},
  doi={10.1109/VTC2024-Fall63153.2024.10757977}
}

@article{shen2010wideband_part1,
  author={Shen, Yuan and Win, Moe Z.},
  title={Fundamental Limits of Wideband Localization---Part {I}: A General Framework},
  journal={IEEE Transactions on Information Theory},
  volume={56},
  number={10},
  pages={4956--4980},
  year={2010},
  doi={10.1109/TIT.2010.2060110}
}

@article{shen2010wideband_part2,
  author={Shen, Yuan and Wymeersch, Henk and Win, Moe Z.},
  title={Fundamental Limits of Wideband Localization---Part {II}: Cooperative Networks},
  journal={IEEE Transactions on Information Theory},
  volume={56},
  number={10},
  pages={4981--5000},
  year={2010},
  doi={10.1109/TIT.2010.2059720}
}

@article{godrich2010mimo_radar,
  author={Godrich, Hana and Haimovich, Alexander M. and Blum, Rick S.},
  title={Target Localization Accuracy Gain in {MIMO} Radar-Based Systems},
  journal={IEEE Transactions on Information Theory},
  volume={56},
  number={6},
  pages={2783--2803},
  year={2010},
  doi={10.1109/TIT.2010.2046246}
}

@ARTICLE{8828030,
  author={Mishra, Kumar Vijay and Bhavani Shankar, M.R. and Koivunen, Visa and Ottersten, Bjorn and Vorobyov, Sergiy A.},
  journal={IEEE Signal Processing Magazine}, 
  title={Toward Millimeter-Wave Joint Radar Communications: A Signal Processing Perspective}, 
  year={2019},
  volume={36},
  number={5},
  pages={100-114},
  doi={10.1109/MSP.2019.2913173}}

@ARTICLE{8642926,
  author={Dokhanchi, Sayed Hossein and Mysore, Bhavani Shankar and Mishra, Kumar Vijay and Ottersten, Björn},
  journal={IEEE Transactions on Aerospace and Electronic Systems}, 
  title={A mmWave Automotive Joint Radar-Communications System}, 
  year={2019},
  volume={55},
  number={3},
  pages={1241-1260},
  doi={10.1109/TAES.2019.2899797}}

@inproceedings{Hong2026IRS,
  author    = {K. Hong and M. Ahmadi and T. Stifter and B. S. M. R. Rao},
  title     = {{Distributed Bistatic Doppler Localization Using a Blender-Based RF Digital Twin for OFDM-Based Wi-Fi Sensing}},
  booktitle = {2026 27th International Radar Symposium (IRS)},
  address   = {Krakow, Poland},
  year      = {2026},
  pages     = {1--6},
  doi       = {10.23919/IRS70539.2026.11549061}
}
